\documentclass[
reprint,
superscriptaddress,
nolongbibliography,
amsmath,
amssymb,
aps,prl]{revtex4-2}
\usepackage{graphicx}
\usepackage{xcolor}
\usepackage[mathlines]{lineno}
\usepackage[
				colorlinks = true,
	            linkcolor = blue,
	            urlcolor  = blue,
	            citecolor = blue,
	            anchorcolor = blue
	            ]{hyperref}
\begin{document}

\title{Experimental observation of exceptional bound states in the continuum}

\author{Shuang Wu}
\affiliation{Institute of Acoustics, School of Physics Science and Engineering, Tongji University, Shanghai 200092, China}
\author{Ruizhi Dong}
\email{ruizhi\_dong@tongji.edu.cn}
\affiliation{Institute of Acoustics, School of Physics Science and Engineering, Tongji University, Shanghai 200092, China}

\author{Nikolay Solodovchenko}
\affiliation{School of Physics and Engineering, ITMO University, St. Petersburg 197101, Russia}
\affiliation{Qingdao Innovation and Development Center, Harbin Engineering University, Qingdao 266000, China}

\author{Dongxing Mao}
\affiliation{Institute of Acoustics, School of Physics Science and Engineering, Tongji University, Shanghai 200092, China}

\author{Andrey Bogdanov}
\email{a.bogdanov@metalab.ifmo.ru}
\affiliation{School of Physics and Engineering, ITMO University, St. Petersburg 197101, Russia}
\affiliation{Qingdao Innovation and Development Center, Harbin Engineering University, Qingdao 266000, China}

\author{Yong Li}
\email{yongli@tongji.edu.cn}
\affiliation{Institute of Acoustics, School of Physics Science and Engineering, Tongji University, Shanghai 200092, China}
\affiliation{School of Physics and Engineering, ITMO University, St. Petersburg 197101, Russia}

\begin{abstract}
We experimentally demonstrate second- and third-order exceptional bound states in the continuum (EP-BICs), formed by the merging of two and three symmetry-protected BICs at an exceptional point (EP). Our passive reciprocal acoustic platform consists of symmetry-protected BIC cavities coupled through an acoustic waveguide and enables independent control of intrinsic loss, radiative loss, and near-field coupling between the cavities. A nonuniformly distributed intrinsic loss provides the non-Hermiticity required for EP formation while preserving decoupling from the waveguide radiation channel. The EP-BICs are probed using both near- and far-field excitation. In the near field, local intracavity excitation directly accesses the symmetry-protected BICs and reveals their spectral evolution. For far-field excitation, we intentionally break the protecting geometrical symmetry, converting the BICs into quasi-BICs and making them accessible through transmission measurements. Extending the system to three cavities with graded intrinsic loss realizes a third-order EP-BIC and yields a larger spectral response to the implemented coupling perturbation. These results establish a passive reciprocal route to higher-order exceptional degeneracies with independent control over intrinsic loss, near-field coupling, and radiative access.
\end{abstract}

\maketitle

Bound states in the continuum (BICs) and exceptional points (EPs) are distinct singular structures in the spectra of open wave systems. A BIC is a nonradiating eigenstate whose frequency lies within the continuous spectrum of propagating environmental waves~\cite{Stillinger1975BIC,Friedrich1985BIC,Hsu2016BICReview,Sadreev2021BICReview,Azzam2021BICReview}. BICs have been realized in photonic, plasmonic, microwave, acoustic, and elastic platforms \cite{Plotnik2011OpticalBIC,Zhen2014TopologicalBIC,Sadrieva2019MultipolarBIC,Huang2021SoundTrapping,Deriy2022AcousticBIC,Krasikova2024AcousticBIC,Huang2025MergingBICs,Liang2026PlasmonicBIC}. Their suppression of radiative leakage underlies high-$Q$ resonances and has enabled lasing, nonlinear frequency conversion, sensing, enhanced wave-matter interaction, and efficient wave confinement \cite{Koshelev2018QuasiBIC,Kodigala2017BICLaser,Yesilkoy2019BICSensing,Kang2023BICApplications}. 


An EP is instead a spectral singularity of a non-Hermitian operator at which at least two eigenvalues and their eigenvectors coalesce \cite{Heiss2012EP,Miri2019EPReview,Zhang2026ExceptionalPoints}. EPs have been extensively explored in parity-time-symmetric systems, but they also occur in passive systems without parity-time symmetry, including photonic, acoustic, and elastic structures \cite{ElGanainy2018NonHermitian,PhysRevX.4.031042,Shi2016AcousticEP,Ding2016MultipleEP,Ding2018AnisotropicEP,Wang2019AcousticEP,Zhu2018TopologicalEP,Huang2024AcousticResonances}. Near an EP of order $N$, a generic perturbation produces an $N$th-root eigenvalue splitting. This nonanalytic response has motivated exceptional-point sensing and the construction of higher-order degeneracies \cite{Wiersig2014EPSensing,Chen2017EPSensing,Hodaei2017HigherOrderEP,Wang2019ArbitraryOrderEP,Mandal2021HigherOrderEP,Fang2021HigherOrderEP}. A larger eigenvalue response does not alone imply greater measurement precision because linewidth, noise, excitation, collection, and estimation also determine sensor performance~\cite{Langbein2018NoPrecision,Lau2018NonHermitianSensing,Wiersig2020EPSensingLimits}.

\begin{figure}
\centering
\includegraphics[width=0.97\linewidth]{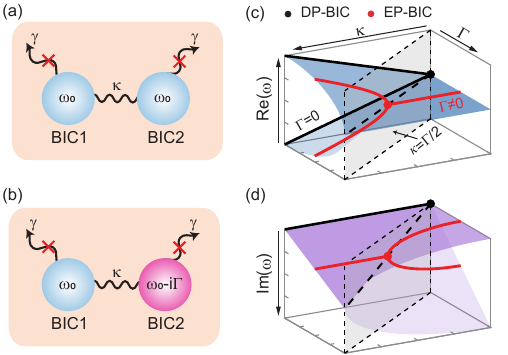}
\caption{Formation of a second-order EP-BIC. (a) Two lossless BICs coupled by $\kappa$ form a diabolic point (DP) as $\kappa\rightarrow0$. The crossed arrows denote a zero radiation rate $\gamma$. (b) One BIC has intrinsic loss $\Gamma$. The modes coalesce at $\kappa=\Gamma/2$ and remain nonradiating. (c,d) Real and imaginary parts of the complex eigenfrequencies as functions of $\kappa$ and $\Gamma$. The black and red paths show the DP-BIC and EP-BIC limits.}
\label{fig1}
\end{figure}

\begin{figure*}
\centering
\includegraphics[width=\linewidth]{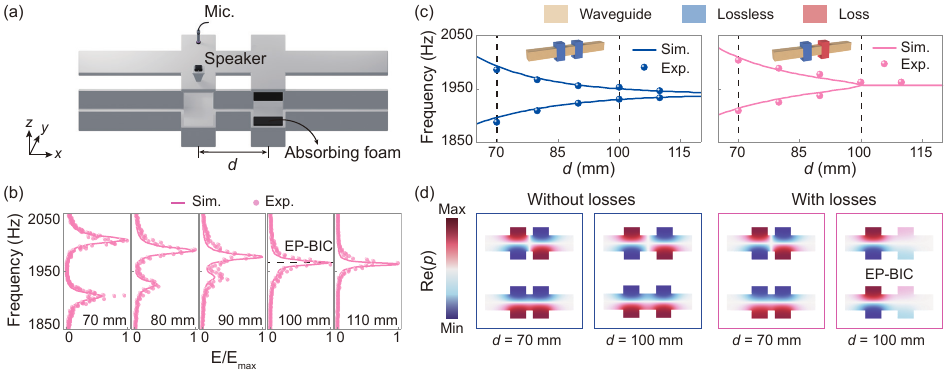}
\caption{Experimental realization of a second-order EP-BIC. (a) Coupled acoustic cavities with separation $d$. A porous insert adds intrinsic loss $\Gamma$ to one cavity. (b) Simulated and measured normalized intracavity spectra for several values of $d$. (c) Simulated eigenfrequencies and measured resonance frequencies for the lossless and lossy systems. (d) Calculated pressure eigenfields at $d=70$ and $100~\mathrm{mm}$ for both systems.}
\label{fig2}
\end{figure*}

Merging BICs at an EP combines nonradiating confinement with a nonanalytic spectral response, offering enhanced eigenvalue sensitivity to perturbations~\cite{CanosValero2025EPBIC}. Related BIC-EP physics was considered earlier in dielectric-waveguide, integrated-photonic, and acoustic settings \cite{Kikkawa2020BICEP,Qin2022EPBIC,Zhou2023ChiralEP}. In a reciprocal passive system, radiation alone cannot generate an EP within an exactly nonradiating BIC subspace. Two identical lossless BICs therefore approach a diabolic point (DP) and form a DP-BIC as their coupling vanishes [Fig.~\ref{fig1}].  For two identical BICs with resonance frequency $\omega_0$, near-field coupling $\kappa$, and intrinsic loss rate $\Gamma$ introduced into one BIC, the effective non-Hermitian Hamiltonian reads
\begin{equation}
H_2=
\begin{bmatrix}
\omega_0 & \kappa\\
\kappa & \omega_0-\rm{i}\Gamma
\end{bmatrix},\ \
\omega_{\pm}=\omega_0-\frac{{\rm i}\Gamma}{2}
\pm\sqrt{\kappa^2-\frac{\Gamma^2}{4}} .
\label{eq:H2}
\end{equation}
The radiative rate remains zero. For $\Gamma=0$, the limit $\kappa\rightarrow0$ gives a DP-BIC with two independent eigenvectors [Fig.~\ref{fig1}(a)]. For $\Gamma>0$, the modes coalesce at $\kappa=\Gamma/2$ into an EP-BIC [Fig.~\ref{fig1}(b)], separating the overcoupled and undercoupled regimes. In the overcoupled regime, $\kappa>\Gamma/2$, the real parts of the eigenfrequencies split while the loss rates coincide. In the undercoupled regime, $\kappa<\Gamma/2$, the real parts coincide while the intrinsic loss rates split [Figs.~\ref{fig1}(c) and \ref{fig1}(d)]. The EP-BIC therefore arises from a precise balance between intrinsic loss and inter-cavity coupling.

Translating this mechanism to optics is experimentally challenging. The scheme proposed in Ref.~\cite{CanosValero2025EPBIC} requires precise control of the subwavelength separation and relative alignment of stacked metasurfaces, together with accurate matching of their resonance frequencies and material losses. A third-order EP-BIC further requires a controlled loss distribution across three coupled layers. Finite-size effects create an additional difficulty because edge radiation converts an ideal BIC into a quasi-BIC \cite{Hoang2025FiniteSize} and can obscure the required balance between intrinsic and radiative loss. Although recent terahertz and elastic experiments have demonstrated connections between BICs and exceptional degeneracies~\cite{Wang2025Photoswitchable,1v48-2t7x,Chen2026PassiveEPBIC}, they have focused on second-order singularities. A controlled experimental realization of a higher-order EP-BIC and its spectral features therefore remain underexplored and technically challenging, requiring precise and independent control of intrinsic loss and inter-cavity coupling.

Here, we propose and experimentally observe second- and third-order EP-BICs in a compact passive reciprocal platform comprising hollow acoustic resonators coupled through a duct. Each cavity supports a symmetry-protected BIC. The system enables independent control over the intrinsic loss of the BICs by introducing absorbing foam into the cavities, the coupling strength between the BICs by varying the cavity separation, and the radiative loss by breaking the cavity symmetry. This independent tunability enables near-field probing of the EP-BICs and far-field readout after controlled symmetry breaking.

The experimental sample represents two mirror-symmetric acoustic cavities coupled evanescently through a single-mode waveguide [Fig.~\ref{fig2}(a)]. The considered cavity modes are odd under the protecting reflection $y\rightarrow-y$ (the local $xz$ mirror plane), whereas the fundamental propagating waveguide mode is even. Their overlap therefore vanishes, producing a symmetry-protected BIC ~\cite{Huang2021SoundTrapping,Deriy2022AcousticBIC,Liu2023MirrorStackingBIC,Yin2025DimensionalTBIC}. The cavity separation $d$ controls the near-field coupling $\kappa$ while preserving symmetry~\cite{Dong2025TopologicalBICSensitivity}. Intrinsic loss is introduced independently by placing porous material symmetrically inside one cavity, which increases $\Gamma$ while preserving the BIC condition. The coupling $\kappa(d)$ is independently calibrated from the frequency splitting of the lossless BIC pair and decreases monotonically with $d$. For more details, see the Supplemental Material, Secs.~I-III~\cite{SuppMat}. It is worth noting that, although both resonances remain nonradiative, their total linewidths and their $Q$ factors remain finite and limited by intrinsic losses.

\begin{figure}[t]
\centering
\includegraphics[width=\linewidth]{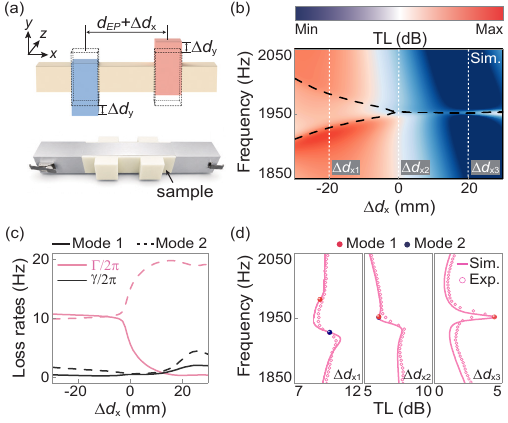}
\caption{Scattering readout of the second-order EP-BIC. (a) Opposite transverse shifts $\pm\Delta d_y$ break the protecting symmetry and introduce radiative loss, converting the BICs into quasi-BICs, while the longitudinal shift $\Delta d_x$ independently tunes the conservative coupling. The lower part shows a schematic of the sample. (b) Calculated transmission-loss (TL) map for $\Delta d_x$. The black dashed lines indicate the calculated eigenfrequencies of the two quasi-BIC modes. (c) Calculated intrinsic ($\Gamma$) and radiative ($\gamma$) loss rates of the two modes for $\Delta d_x$. (d) Simulated and measured TL spectra for $\Delta d_x=-20$, $0$, and $20$ mm. In all panels, $\Delta d_y$ is fixed at $1~\mathrm{mm}$.}
\label{fig3}
\end{figure}

Because a genuine BIC cannot be excited through the propagating waveguide channel, we probe the modes locally using a source and microphone placed inside the cavity region [Fig.~\ref{fig2}(a)]. The measured and simulated spectra are shown in Fig.~\ref{fig2}(b).  The experimental and simulation details are given in the Supplemental Material, Secs.~I and III~\cite{SuppMat}. Figure~\ref{fig2}(c) compares the measured resonance frequencies with the numerically calculated eigenfrequencies. Without additional loss, the two branches gradually approach each other as $d$ increases, as expected when the coupling between two lossless BICs decreases. The modes remain distinct symmetric and antisymmetric combinations of the individual cavity states [Fig.~\ref{fig2}(d)] and retain two independent eigenvectors, corresponding to the DP-BIC limit of Eq.~\eqref{eq:H2}.

Introducing intrinsic loss into one cavity qualitatively changes the branch evolution. The two measured resonance frequencies merge near $d_{\rm EP}\approx100~\mathrm{mm}$ at $f\simeq1.96~\mathrm{kHz}$ [Fig.~\ref{fig2}(c)], where the independently calibrated coupling satisfies $\kappa\simeq\Gamma/2$. At the same separation, the two calculated eigenfields become almost identical [Fig.~\ref{fig2}(d)], demonstrating eigenvector coalescence in addition to eigenvalue coalescence. The evolution on either side of $d_{\rm EP}$ is consistent with the square-root dispersion of Eq.~\eqref{eq:H2}, with frequency splitting for $\kappa>\Gamma/2$ and nearly coincident real parts for $\kappa<\Gamma/2$.

A genuine symmetry-protected BIC is inaccessible to far-field excitation. We therefore shift the two cavities transversely in opposite directions by $\pm\Delta d_y$ [Fig.~\ref{fig3}(a)], breaking the protecting mirror symmetry and converting the BICs into radiative quasi-BICs~\cite{qj87-5xz9, Koshelev2018QuasiBIC}. Independently, the longitudinal displacement $\Delta d_x$ tunes the inter-cavity coupling and moves the system away from the EP condition~\cite{Sakotic2023NonHermitianBIC}, thereby separating modal tuning from radiative readout. The symmetry breaking introduces a finite radiative loss $\gamma$ and lifts the exact EP-BIC condition.  Figure~\ref{fig3}(b) shows the calculated transmission-loss (TL) map versus frequency and $\Delta d_x$. Because the symmetry breaking is weak, the quasi-BIC branches remain close to the underlying EP-BIC evolution. Figure~\ref{fig3}(c) shows the intrinsic $\Gamma$ and radiative $\gamma$ loss rates, governed mainly by the modal overlap with the lossy cavity and coupling to the waveguide, respectively.  The loss decomposition is described in the Supplemental Material,  Sec.~IV~\cite{SuppMat}.

We next experimentally measure the TL spectra in three representative regimes: (i) the overcoupled regime at $\Delta d_x=-20~\mathrm{mm}$, (ii) the EP-BIC condition at $\Delta d_x=0$, and (iii) the undercoupled regime at $\Delta d_x=20~\mathrm{mm}$. The experimental sample is shown in Fig.~\ref{fig3}(a). Figure~\ref{fig3}(d) compares the measured and calculated TL spectra. The simulations include propagation loss and experimentally characterized background dissipation and show excellent agreement with the measurements in all three regimes, reproducing both the resonance positions and their spectral evolution as the coupling is varied. Transmission is extracted from the upstream and downstream pressure fields using the four-microphone transfer-matrix method \cite{ASTM}, with experimental details given in the Supplemental Material, Sec.~IV~\cite{SuppMat}. The red and blue points indicate the calculated positions of the two quasi-BICs, which approach one another most closely near the underlying EP-BIC condition at $\Delta d_x=0$. Therefore, we demonstrate that controlled symmetry breaking converts the genuine EP-BIC into weakly radiative quasi-BICs, thereby allowing the underlying EP-BIC physics to be probed from the far field through transmission spectra.

\begin{figure*}
\centering
\includegraphics[width=\linewidth]{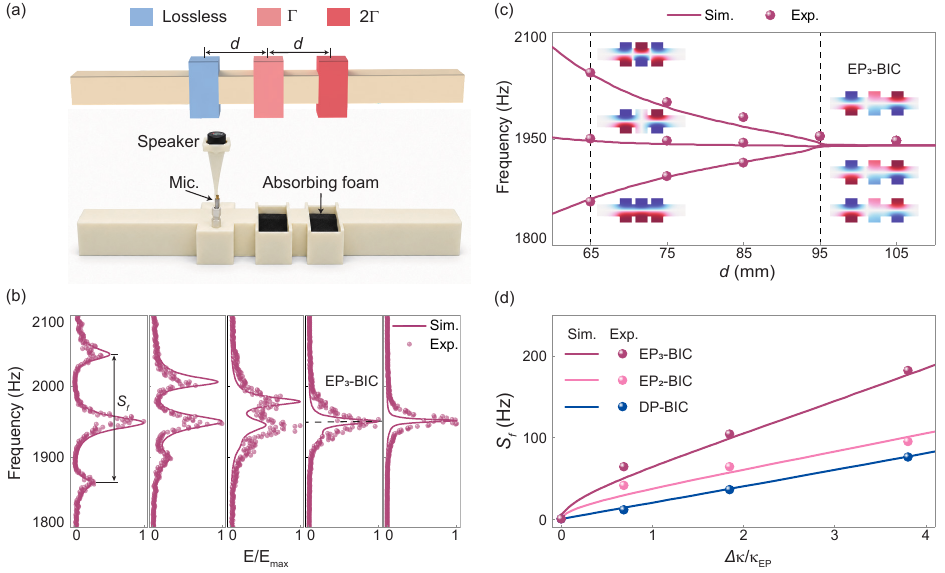}
\caption{Third-order EP-BIC and spectral response.
(a) Three coupled BIC cavities with intrinsic loss rates $0$, $\Gamma$, and $2\Gamma$. The lower part shows a schematic of the experimental setup.
(b) Simulated and measured intracavity spectra as the separation $d$ is varied, showing mode coalescence near $d=95~\mathrm{mm}$.
(c) Corresponding eigenfrequencies and representative pressure fields.
(d) Change in spectral span $S_f$ under a coupling perturbation for the DP-BIC, EP$_2$-BIC, and EP$_3$-BIC. For the DP-BIC data, $\kappa_{\rm EP}$ in the horizontal-axis normalization is taken as the EP$_2$ coupling strength, solely as a common reference scale. In all panels, curves denote simulations and symbols denote experiments.}\label{fig4}
\end{figure*}

The concept of forming an EP within the BIC subspace through a controlled balance of coupling and intrinsic loss can be generalized to a third-order EP~\cite{CanosValero2025EPBIC}. For three identical BICs with equal nearest-neighbor coupling $\kappa$ and intrinsic losses $0$, $\Gamma$, and $2\Gamma$, the effective Hamiltonian is
\begin{equation}
H_3=
\begin{bmatrix}
\omega_0 & \kappa & 0\\
\kappa & \omega_0-{\rm i}\Gamma & \kappa\\
0 & \kappa & \omega_0-2{\rm i}\Gamma
\end{bmatrix}.
\label{eq:H3}
\end{equation}
Its eigenfrequencies are $\omega_0-{\rm i}\Gamma$ and $\omega_0-{\rm i}\Gamma\pm\sqrt{2\kappa^2-\Gamma^2}$. At $\kappa=\Gamma/\sqrt{2}$, all three eigenfrequencies coalesce and the eigenvectors collapse onto $[1,-\rm{i}\sqrt{2},-1]^T$, forming a third-order Jordan block \cite{Hodaei2017HigherOrderEP,Wang2019ArbitraryOrderEP,Mandal2021HigherOrderEP}. 

To realize this condition experimentally and observe a third-order EP-BIC, we extend the two-cavity configuration to three identical symmetry-protected acoustic cavities with a common nearest-neighbor separation $d$ [Fig.~\ref{fig4}(a)], similarly to Ref.~\cite{Song2026TripleResonatorBIC}. Absorbing foam is introduced into the central and one outer cavity, with the amount of foam adjusted to emulate the required intrinsic loss rates $\Gamma$ and $2\Gamma$, respectively, while the remaining cavity is kept nominally lossless. Varying $d$ tunes the two nearest-neighbor couplings simultaneously. The observation scheme is identical to that used for the second-order EP-BIC in Fig.~\ref{fig2}. See the Supplemental Material, Sec.~V for more details~\cite{SuppMat}.

Figure~\ref{fig4}(b) shows the measured and simulated intracavity spectra for several values of $d$. At a small separation, three distinct resonances are clearly resolved. As $d$ increases and the coupling decreases, their spectral span $S_f$ continuously shrinks, and the three resonances coalesce into the third-order EP-BIC near $d\simeq95~\mathrm{mm}$. The measured spectra closely follow the simulations throughout this evolution. Figure~\ref{fig4}(c) compares the numerically calculated eigenfrequencies with the resonance frequencies extracted from the measured spectra. The calculated eigenfrequencies and measured resonance frequencies agree well over the entire range of $d$ and converge near $d\simeq95~\mathrm{mm}$. The calculated pressure distributions in the inset show that the three initially distinct modes coalesce into a common modal profile at the third-order EP-BIC.

Having established the third-order EP-BIC, we compare the spectral response of the different degeneracies. As a regular reference, we use the DP-BIC with two independent eigenvectors. In all cases, the perturbation modifies only the coupling $\kappa$, thereby isolating the eigenvalue response from radiative effects. Figure~\ref{fig4}(d) shows the resulting spectral span $S_f=f_{\max}-f_{\min}$. For the DP-BIC, the degeneracy is lifted linearly, $S_f=|\Delta\kappa|/\pi$. In contrast, the EP-BICs exhibit a nonanalytic response. For the second-order EP-BIC, substituting $\kappa=\kappa_{\rm EP}+\Delta\kappa$, with $\kappa_{\rm EP}=\Gamma/2$, into Eq.~\eqref{eq:H2} gives $S_f=\sqrt{2\kappa_{\rm EP}\Delta\kappa+\Delta\kappa^2}/\pi$, and therefore $S_f\propto|\Delta\kappa|^{1/2}$ sufficiently close to the EP. For the third-order EP-BIC, with $\kappa_{\rm EP}=\Gamma/\sqrt{2}$, Eq.~\eqref{eq:H3} yields $S_f=\sqrt{4\kappa_{\rm EP}\Delta\kappa+2\Delta\kappa^2}/\pi$, which also scales as $|\Delta\kappa|^{1/2}$. The enhanced nonanalytic response occurs for $\Delta\kappa/\kappa_{\rm EP}\lesssim2$, whereas for larger perturbations the spectral splitting crosses over to an approximately linear dependence on $\Delta\kappa$, see the Supplemental Material, Sec.~V~\cite{SuppMat}. Experiment and simulation reproduce this nonanalytic response, with the EP$_3$-BIC exhibiting the largest spectral splitting. The square-root response of the EP$_3$-BIC deserves emphasis. A generic perturbation of a third-order Jordan block can produce the characteristic cubic-root eigenvalue splitting. The present perturbation, however, changes the two nearest-neighbor couplings simultaneously and therefore follows a constrained direction in parameter space. Along this direction, one eigenvalue remains pinned at $\omega_0-{\rm i}\Gamma$, while the other two split with a square-root dependence.  Because the coupling perturbation preserves the BIC condition, the larger spectral splitting of the EP$_3$-BIC relative to the EP$_2$-BIC and DP-BIC originates from their different eigenvalue responses along the chosen perturbation direction, rather than from radiative effects. However, a larger spectral splitting does not necessarily imply improved sensing precision, which additionally depends on linewidth, noise, excitation and detection efficiencies, and the estimation protocol~\cite{Lau2018NonHermitianSensing,Wiersig2020EPSensingLimits}.

We have experimentally realized second- and third-order exceptional points within an acoustic BIC subspace. Controlled intrinsic dissipation provides the non-Hermiticity required for exceptional degeneracies while preserving decoupling from the propagating waveguide, and the cavity separation independently tunes the conservative inter-cavity coupling. Local spectral measurements together with numerical eigenmode analysis reveal eigenvalue and eigenvector coalescence, while the protecting symmetry ensures that the coalesced modes retain their nonradiating BIC character. By weakly breaking the protecting symmetry, the second-order EP-BIC is converted into a pair of quasi-EP-BICs and becomes directly accessible in transmission. Finally, the three-cavity system realizes a third-order EP-BIC and exhibits an enhanced spectral response to the coupling perturbation. These results establish a general route to engineering higher-order exceptional degeneracies in nonradiating subspaces and provide a platform for independently controlling non-Hermiticity, modal coupling, and radiative accessibility.

\textit{Acknowledgments--}This work was supported by the Russian Science Foundation (Grant No. 25-79-31027).

\textit{Data availability--}The data that support the findings of this Letter are not publicly available. The data are available from the authors upon reasonable request.

\bibliographystyle{apsrev4-2}
\bibliography{references-v2}

@misc{SuppMat,
title = {See {Supplemental Material} at [url will be inserted by publisher] for the formulation of the acoustic eigenproblem, derivation of the effective coupled-mode Hamiltonians for second- and third-order {EP-BICs}, details of the numerical simulations and experimental characterization of second-order {EP-BICs}, extraction of intrinsic and radiative decay rates, transmission measurements, and realization and spectral analysis of a three-cavity system supporting a third-order {EP-BIC}}
}

@article{Kikkawa2020BICEP,
  author  = {Kikkawa, Ryo and Nishida, Munehiro and Kadoya, Yutaka},
  title   = {Bound states in the continuum and exceptional points in dielectric waveguide equipped with a metal grating},
  journal = {New J. Phys.},
  volume  = {22},
  number  = {7},
  pages   = {073029},
  year    = {2020},
  doi     = {10.1088/1367-2630/ab97e9},
  url     = {https://doi.org/10.1088/1367-2630/ab97e9}
}

@article{Qin2022EPBIC,
  author  = {Qin, Haoye and Shi, Xiaodong and Ou, Haiyan},
  title   = {Exceptional points at bound states in the continuum in photonic integrated circuits},
  journal = {Nanophotonics},
  volume  = {11},
  number  = {21},
  pages   = {4909--4917},
  year    = {2022},
  doi     = {10.1515/nanoph-2022-0420},
  url     = {https://doi.org/10.1515/nanoph-2022-0420}
}

@article{Stillinger1975BIC,
  author = {Stillinger, Frank H. and Herrick, David R.},
  title = {{Bound states in the continuum}},
  journal = {Phys. Rev. A},
  year = {1975},
  volume = {11},
  number = {2},
  pages = {446--454},
  doi = {10.1103/PhysRevA.11.446},
  url = {https://doi.org/10.1103/PhysRevA.11.446},
}

@article{Friedrich1985BIC,
  author = {Friedrich, H. and Wintgen, D.},
  title = {{Interfering resonances and bound states in the continuum}},
  journal = {Phys. Rev. A},
  year = {1985},
  volume = {32},
  number = {6},
  pages = {3231--3242},
  doi = {10.1103/PhysRevA.32.3231},
  url = {https://doi.org/10.1103/PhysRevA.32.3231},
}

@article{Hsu2016BICReview,
  author = {Hsu, Chia Wei and Zhen, Bo and Stone, A. Douglas and Joannopoulos, John D. and Solja{\v{c}}i{\'c}, Marin},
  title = {{Bound states in the continuum}},
  journal = {Nat. Rev. Mater.},
  year = {2016},
  volume = {1},
  number = {9},
  pages = {16048},
  doi = {10.1038/natrevmats.2016.48},
  url = {https://doi.org/10.1038/natrevmats.2016.48},
}

@article{Sadreev2021BICReview,
  author = {Sadreev, Almas F},
  title = {{Interference traps waves in an open system: bound states in the continuum}},
  journal = {Rep. Prog. Phys.},
  year = {2021},
  volume = {84},
  number = {5},
  pages = {055901},
  doi = {10.1088/1361-6633/abefb9},
  url = {https://doi.org/10.1088/1361-6633/abefb9},
}

@article{Koshelev2018QuasiBIC,
  author = {Koshelev, Kirill and Lepeshov, Sergey and Liu, Mingkai and Bogdanov, Andrey and Kivshar, Yuri},
  title = {{Asymmetric Metasurfaces with High-{$Q$} Resonances Governed by Bound States in the Continuum}},
  journal = {Phys. Rev. Lett.},
  year = {2018},
  volume = {121},
  number = {19},
  pages = {193903},
  doi = {10.1103/PhysRevLett.121.193903},
  url = {https://doi.org/10.1103/PhysRevLett.121.193903},
}

@article{Kodigala2017BICLaser,
  author = {Kodigala, Ashok and Lepetit, Thomas and Gu, Qing and Bahari, Babak and Fainman, Yeshaiahu and Kant{\'e}, Boubacar},
  title = {{Lasing action from photonic bound states in continuum}},
  journal = {Nature},
  year = {2017},
  volume = {541},
  number = {7636},
  pages = {196--199},
  doi = {10.1038/nature20799},
  url = {https://doi.org/10.1038/nature20799},
}

@article{Zhen2014TopologicalBIC,
  author = {Zhen, Bo and Hsu, Chia Wei and Lu, Ling and Stone, A. Douglas and Solja{\v{c}}i{\'c}, Marin},
  title = {{Topological Nature of Optical Bound States in the Continuum}},
  journal = {Phys. Rev. Lett.},
  year = {2014},
  volume = {113},
  number = {25},
  pages = {257401},
  doi = {10.1103/PhysRevLett.113.257401},
  url = {https://doi.org/10.1103/PhysRevLett.113.257401},
}

@article{Sadrieva2019MultipolarBIC,
  author = {Sadrieva, Zarina and Frizyuk, Kristina and Petrov, Mihail and Kivshar, Yuri and Bogdanov, Andrey},
  title = {{Multipolar Origin of Bound States in the Continuum}},
  journal = {Phys. Rev. B},
  year = {2019},
  volume = {100},
  number = {11},
  pages = {115303},
  doi = {10.1103/PhysRevB.100.115303},
  url = {https://doi.org/10.1103/PhysRevB.100.115303},
}

@article{Miri2019EPReview,
  author = {Miri, Mohammad-Ali and Al{\`u}, Andrea},
  title = {{Exceptional points in optics and photonics}},
  journal = {Science},
  year = {2019},
  volume = {363},
  number = {6422},
  pages = {eaar7709},
  doi = {10.1126/science.aar7709},
  url = {https://doi.org/10.1126/science.aar7709},
}

@article{ElGanainy2018NonHermitian,
  author = {El-Ganainy, Ramy and Makris, Konstantinos G. and Khajavikhan, Mercedeh and Musslimani, Ziad H. and Rotter, Stefan and Christodoulides, Demetrios N.},
  title = {{Non-Hermitian physics and {PT} symmetry}},
  journal = {Nat. Phys.},
  year = {2018},
  volume = {14},
  number = {1},
  pages = {11--19},
  doi = {10.1038/nphys4323},
  url = {https://doi.org/10.1038/nphys4323},
}

@article{CanosValero2025EPBIC,
  author = {Can{\'o}s Valero, Adri{\`a} and Sztranyovszky, Zoltan and Muljarov, Egor A. and Bogdanov, Andrey and Weiss, Thomas},
  title = {{Exceptional Bound States in the Continuum}},
  journal = {Phys. Rev. Lett.},
  year = {2025},
  volume = {134},
  number = {10},
  pages = {103802},
  doi = {10.1103/PhysRevLett.134.103802},
  url = {https://doi.org/10.1103/PhysRevLett.134.103802},
}

@article{Sakotic2023NonHermitianBIC,
  author = {Sakotic, Zarko and Stankovic, Predrag and Bengin, Vesna and Krasnok, Alex and Al{\'u}, Andrea and Jankovic, Nikolina},
  title = {{Non-Hermitian Control of Topological Scattering Singularities Emerging from Bound States in the Continuum}},
  journal = {Laser Photonics Rev.},
  year = {2023},
  volume = {17},
  number = {6},
  pages = {2200308},
  doi = {10.1002/lpor.202200308},
  url = {https://doi.org/10.1002/lpor.202200308},
}

@article{1v48-2t7x,
  author = {Cao, Liyun and Assouar, Badreddine},
  title = {Asymmetric Elastic Bound State in the Continuum by an Exceptional Point},
  journal = {Phys. Rev. Lett.},
  year = {2025},
  volume = {135},
  issue = {24},
  pages = {246301},
  numpages = {8},
  month = {Dec},
  publisher = {American Physical Society},
  doi = {10.1103/1v48-2t7x},
  url = {https://link.aps.org/doi/10.1103/1v48-2t7x},
}

@article{Huang2021SoundTrapping,
  author = {Huang, Lujun and Chiang, Yan Kei and Huang, Sibo and Shen, Chen and Deng, Fu and Cheng, Yi and Jia, Bin and Li, Yong and Powell, David A. and Miroshnichenko, Andrey E.},
  title = {{Sound trapping in an open resonator}},
  journal = {Nat. Commun.},
  year = {2021},
  volume = {12},
  number = {1},
  pages = {4819},
  doi = {10.1038/s41467-021-25130-4},
  url = {https://doi.org/10.1038/s41467-021-25130-4},
}

@article{Deriy2022AcousticBIC,
  author = {Deriy, Ilya and Toftul, Ivan and Petrov, Mihail and Bogdanov, Andrey},
  title = {{Bound States in the Continuum in Compact Acoustic Resonators}},
  journal = {Phys. Rev. Lett.},
  year = {2022},
  volume = {128},
  number = {8},
  pages = {084301},
  doi = {10.1103/PhysRevLett.128.084301},
  url = {https://doi.org/10.1103/PhysRevLett.128.084301},
}

@article{Liu2023MirrorStackingBIC,
  author = {Liu, Luohong and Li, Tianzi and Zhang, Qicheng and Xiao, Meng and Qiu, Chunyin},
  title = {{Universal Mirror-Stacking Approach for Constructing Topological Bound States in the Continuum}},
  journal = {Phys. Rev. Lett.},
  year = {2023},
  volume = {130},
  number = {10},
  pages = {106301},
  doi = {10.1103/PhysRevLett.130.106301},
  url = {https://doi.org/10.1103/PhysRevLett.130.106301},
}

@article{Yin2025DimensionalTBIC,
  author = {Yin, Shunda and Wang, Zhenyu and Ye, Liping and He, Hailong and Ke, Manzhu and Deng, Weiyin and Lu, Jiuyang and Liu, Zhengyou},
  title = {{Dimensional Hierarchy of Topological Bound States in the Continuum}},
  journal = {Phys. Rev. Lett.},
  year = {2025},
  volume = {135},
  number = {12},
  pages = {126602},
  doi = {10.1103/lq9m-nngh},
  url = {https://doi.org/10.1103/lq9m-nngh},
}

@article{PhysRevX.4.031042,
  author = {Zhu, Xuefeng and Ramezani, Hamidreza and Shi, Chengzhi and Zhu, Jie and Zhang, Xiang},
  title = {$\mathcal{P}\mathcal{T}$-Symmetric Acoustics},
  journal = {Phys. Rev. X},
  year = {2014},
  volume = {4},
  issue = {3},
  pages = {031042},
  numpages = {7},
  month = {Sep},
  publisher = {American Physical Society},
  doi = {10.1103/PhysRevX.4.031042},
  url = {https://link.aps.org/doi/10.1103/PhysRevX.4.031042},
}

@article{Shi2016AcousticEP,
  author = {Shi, Chengzhi and Dubois, Marc and Chen, Yun and Cheng, Lei and Ramezani, Hamidreza and Wang, Yuan and Zhang, Xiang},
  title = {{Accessing the exceptional points of parity-time symmetric acoustics}},
  journal = {Nat. Commun.},
  year = {2016},
  volume = {7},
  number = {1},
  pages = {11110},
  doi = {10.1038/ncomms11110},
  url = {https://doi.org/10.1038/ncomms11110},
}

@article{Ding2016MultipleEP,
  author = {Ding, Kun and Ma, Guancong and Xiao, Meng and Zhang, Z. Q. and Chan, C. T.},
  title = {{Emergence, Coalescence, and Topological Properties of Multiple Exceptional Points and Their Experimental Realization}},
  journal = {Phys. Rev. X},
  year = {2016},
  volume = {6},
  number = {2},
  pages = {021007},
  doi = {10.1103/PhysRevX.6.021007},
  url = {https://doi.org/10.1103/PhysRevX.6.021007},
}

@article{Ding2018AnisotropicEP,
  author = {Ding, Kun and Ma, Guancong and Zhang, Z. Q. and Chan, C. T.},
  title = {{Experimental Demonstration of an Anisotropic Exceptional Point}},
  journal = {Phys. Rev. Lett.},
  year = {2018},
  volume = {121},
  number = {8},
  pages = {085702},
  doi = {10.1103/PhysRevLett.121.085702},
  url = {https://doi.org/10.1103/PhysRevLett.121.085702},
}

@article{Wiersig2014EPSensing,
  author = {Wiersig, Jan},
  title = {{Enhancing the Sensitivity of Frequency and Energy Splitting Detection by Using Exceptional Points: Application to Microcavity Sensors for Single-Particle Detection}},
  journal = {Phys. Rev. Lett.},
  year = {2014},
  volume = {112},
  number = {20},
  pages = {203901},
  doi = {10.1103/PhysRevLett.112.203901},
  url = {https://doi.org/10.1103/PhysRevLett.112.203901},
}

@article{Chen2017EPSensing,
  author = {Chen, Weijian and Kaya {\"O}zdemir, {\c{S}}ahin and Zhao, Guangming and Wiersig, Jan and Yang, Lan},
  title = {{Exceptional points enhance sensing in an optical microcavity}},
  journal = {Nature},
  year = {2017},
  volume = {548},
  number = {7666},
  pages = {192--196},
  doi = {10.1038/nature23281},
  url = {https://doi.org/10.1038/nature23281},
}

@article{Hodaei2017HigherOrderEP,
  author = {Hodaei, Hossein and Hassan, Absar U. and Wittek, Steffen and Garcia-Gracia, Hipolito and El-Ganainy, Ramy and Christodoulides, Demetrios N. and Khajavikhan, Mercedeh},
  title = {{Enhanced sensitivity at higher-order exceptional points}},
  journal = {Nature},
  year = {2017},
  volume = {548},
  number = {7666},
  pages = {187--191},
  doi = {10.1038/nature23280},
  url = {https://doi.org/10.1038/nature23280},
}

@article{Wang2019ArbitraryOrderEP,
  author = {Wang, Shubo and Hou, Bo and Lu, Weixin and Chen, Yuntian and Zhang, Z. Q. and Chan, C. T.},
  title = {{Arbitrary order exceptional point induced by photonic spin--orbit interaction in coupled resonators}},
  journal = {Nat. Commun.},
  year = {2019},
  volume = {10},
  number = {1},
  pages = {832},
  doi = {10.1038/s41467-019-08826-6},
  url = {https://doi.org/10.1038/s41467-019-08826-6},
}

@article{Mandal2021HigherOrderEP,
  author = {Mandal, Ipsita and Bergholtz, Emil J.},
  title = {{Symmetry and Higher-Order Exceptional Points}},
  journal = {Phys. Rev. Lett.},
  year = {2021},
  volume = {127},
  number = {18},
  pages = {186601},
  doi = {10.1103/PhysRevLett.127.186601},
  url = {https://doi.org/10.1103/PhysRevLett.127.186601},
}

@article{Dong2025TopologicalBICSensitivity,
  author = {Dong, Ruizhi and Zhu, Yihuan and Mao, Dongxing and Wang, Xu and Li, Yong},
  title = {{Observation of Extreme Anisotropic Sensitivity at Topological Bound States in the Continuum}},
  journal = {Phys. Rev. Lett.},
  year = {2025},
  volume = {134},
  number = {20},
  pages = {206601},
  doi = {10.1103/PhysRevLett.134.206601},
  url = {https://doi.org/10.1103/PhysRevLett.134.206601},
}

@article{Lau2018NonHermitianSensing,
  author = {Lau, Hoi-Kwan and Clerk, Aashish A.},
  title = {{Fundamental limits and non-reciprocal approaches in non-Hermitian quantum sensing}},
  journal = {Nat. Commun.},
  year = {2018},
  volume = {9},
  number = {1},
  pages = {4320},
  doi = {10.1038/s41467-018-06477-7},
  url = {https://doi.org/10.1038/s41467-018-06477-7},
}

@article{Wiersig2020EPSensingLimits,
  author = {Wiersig, Jan},
  title = {{Prospects and fundamental limits in exceptional point-based sensing}},
  journal = {Nat. Commun.},
  year = {2020},
  volume = {11},
  number = {1},
  pages = {2454},
  doi = {10.1038/s41467-020-16373-8},
  url = {https://doi.org/10.1038/s41467-020-16373-8},
}

@misc{ASTM,
  author = {{ASTM International}},
  title = {Standard Test Method for Normal Incidence Determination of Porous Material Acoustical Properties Based on the Transfer Matrix Method},
  number = {ASTM E2611-24},
  year = {2024},
  doi = {10.1520/E2611-24},
  url = {https://doi.org/10.1520/E2611-24},
}

@article{Zhou2023ChiralEP,
  author = {Zhou, Zhiling and Jia, Bin and Wang, Nengyin and Wang, Xu and Li, Yong},
  title = {Observation of Perfectly-Chiral Exceptional Point via Bound State in the Continuum},
  journal = {Phys. Rev. Lett.},
  year = {2023},
  volume = {130},
  number = {11},
  pages = {116101},
  doi = {10.1103/PhysRevLett.130.116101},
  url = {https://doi.org/10.1103/PhysRevLett.130.116101},
}

@article{Krasikova2024AcousticBIC,
  author = {Krasikova, Mariia and Kronowetter, Felix and Krasikov, Sergey and Kuzmin, Mikhail and Maeder, Marcus and Yang, Tao and Melnikov, Anton and Marburg, Steffen and Bogdanov, Andrey},
  title = {Acoustic bound states in the continuum in coupled {Helmholtz} resonators},
  journal = {Phys. Rev. Appl.},
  year = {2024},
  volume = {22},
  number = {2},
  pages = {024045},
  doi = {10.1103/PhysRevApplied.22.024045},
  url = {https://doi.org/10.1103/PhysRevApplied.22.024045},
}

@article{Wang2025Photoswitchable,
  author = {Wang, Lei and Liu, Hang and Liu, Junwei and Liu, Aoxuan and Huang, Jialiang and Li, Qiannan and Dai, Hui and Zhang, Caihong and Wu, Jingbo and Fan, Kebin and Wang, Huabing and Jin, Biaobing and Chen, Jian and Wu, Peiheng},
  title = {Photoswitchable exceptional points derived from bound states in the continuum},
  journal = {Light Sci. Appl.},
  year = {2025},
  volume = {14},
  pages = {377},
  doi = {10.1038/s41377-025-02036-0},
  url = {https://doi.org/10.1038/s41377-025-02036-0},
}

@article{Chen2026PassiveEPBIC,
  author = {Chen, Chen and Liu, Ziyi and Li, Jianfei and Zhou, Junhan and Jiang, Yuanhang and Wang, Ying and Zhou, Zhongxiang and Liu, Hongyu and Yao, Jingfeng and Liu, Gui-Geng and Yuan, Chengxun},
  title = {Exceptional Bound States in the Continuum in a Passive Terahertz Metasurface},
  journal = {Research Square},
  year = {2026},
  doi = {10.21203/rs.3.rs-9232537/v1},
  note = {Preprint},
  url = {https://doi.org/10.21203/rs.3.rs-9232537/v1},
}

@article{Langbein2018NoPrecision,
  author = {Langbein, Wolfgang},
  title = {No exceptional precision of exceptional-point sensors},
  journal = {Phys. Rev. A},
  year = {2018},
  volume = {98},
  number = {2},
  pages = {023805},
  doi = {10.1103/PhysRevA.98.023805},
  url = {https://doi.org/10.1103/PhysRevA.98.023805},
}

@article{Azzam2021BICReview,
  author = {Azzam, Shaimaa I. and Kildishev, Alexander V.},
  title = {Photonic Bound States in the Continuum: From Basics to Applications},
  journal = {Adv. Opt. Mater.},
  year = {2021},
  volume = {9},
  number = {1},
  pages = {2001469},
  doi = {10.1002/adom.202001469},
  url = {https://doi.org/10.1002/adom.202001469},
}

@article{Plotnik2011OpticalBIC,
  author = {Plotnik, Yonatan and Peleg, Or and Dreisow, Felix and Heinrich, Matthias and Nolte, Stefan and Szameit, Alexander and Segev, Mordechai},
  title = {Experimental Observation of Optical Bound States in the Continuum},
  journal = {Phys. Rev. Lett.},
  year = {2011},
  volume = {107},
  number = {18},
  pages = {183901},
  doi = {10.1103/PhysRevLett.107.183901},
  url = {https://doi.org/10.1103/PhysRevLett.107.183901},
}

@article{Yesilkoy2019BICSensing,
  author = {Yesilkoy, Filiz and Arvelo, Eduardo R. and Jahani, Yasaman and Liu, Mingkai and Tittl, Andreas and Cevher, Volkan and Kivshar, Yuri and Altug, Hatice},
  title = {Ultrasensitive Hyperspectral Imaging and Biodetection Enabled by Dielectric Metasurfaces},
  journal = {Nat. Photonics},
  year = {2019},
  volume = {13},
  number = {6},
  pages = {390--396},
  doi = {10.1038/s41566-019-0394-6},
  url = {https://doi.org/10.1038/s41566-019-0394-6},
}

@article{Heiss2012EP,
  author = {Heiss, W. D.},
  title = {The Physics of Exceptional Points},
  journal = {J. Phys. A: Math. Theor.},
  year = {2012},
  volume = {45},
  number = {44},
  pages = {444016},
  doi = {10.1088/1751-8113/45/44/444016},
  url = {https://doi.org/10.1088/1751-8113/45/44/444016},
}

@article{Hoang2025FiniteSize,
  author = {Hoang, Thanh Xuan and Leykam, Daniel and Chu, Hong-Son and Png, Ching Eng and Garc{\'i}a-Vidal, Francisco J. and Kivshar, Yuri S.},
  title = {Collective Nature of High-{$Q$} Resonances in Finite-Size Photonic Metastructures},
  journal = {Phys. Rev. Research},
  year = {2025},
  volume = {7},
  number = {1},
  pages = {013316},
  doi = {10.1103/PhysRevResearch.7.013316},
  url = {https://doi.org/10.1103/PhysRevResearch.7.013316},
}

@article{Kang2023BICApplications,
  author = {Kang, Meng and Liu, Tao and Chan, C. T. and Xiao, Meng},
  title = {{Applications of bound states in the continuum in photonics}},
  journal = {Nat. Rev. Phys.},
  year = {2023},
  volume = {5},
  number = {11},
  pages = {659--678},
  doi = {10.1038/s42254-023-00642-8},
  url = {https://doi.org/10.1038/s42254-023-00642-8},
}

@article{Liang2026PlasmonicBIC,
  author = {Liang, Yao and Lepeshov, Sergei and Koshelev, Kirill and Bogdanov, Andrey and Tsai, Din Ping and Kivshar, Yuri S.},
  title = {{Bound states in the continuum in plasmonic structures}},
  journal = {Rep. Prog. Phys.},
  year = {2026},
  doi = {10.1088/1361-6633/ae8311},
  note = {Online published 26 June 2026},
  url = {https://doi.org/10.1088/1361-6633/ae8311},
}

@article{Zhang2026ExceptionalPoints,
  author        = {Fan Zhang and Nikolay Solodovchenko and Dmitrii N. Maksimov and Xuchen Wang and Mingzhao Song and Filippo Capolino and C. T. Chan and Andrey Bogdanov},
  title         = {Exceptional Points in Photonics: From Non-Hermitian Physics to Applications},
  year          = {2026},
  pages        = {2609.01239},
  journal = {arXiv},
  doi           = {10.48550/arXiv.2609.01239},
  url           = {https://arxiv.org/abs/2609.01239}
}

@article{Fang2021HigherOrderEP,
  author  = {Xinsheng Fang and Nikhil J. R. K. Gerard and Zhiling Zhou and Hua Ding and Nengyin Wang and Bin Jia and Yuanchen Deng and Xu Wang and Yun Jing and Yong Li},
  title   = {Observation of higher-order exceptional points in a non-local acoustic metagrating},
  journal = {Commun. Phys.},
  volume  = {4},
  pages   = {271},
  year    = {2021},
  doi     = {10.1038/s42005-021-00779-x}
}

@article{Huang2024AcousticResonances,
  author  = {Lujun Huang and Sibo Huang and Chen Shen and Simon Yves and Artem S. Pilipchuk and Xiang Ni and Seunghwi Kim and Yan Kei Chiang and David A. Powell and Jie Zhu and Ya Cheng and Yong Li and Almas F. Sadreev and Andrea Al{\`u} and Andrey E. Miroshnichenko},
  title   = {Acoustic resonances in non-Hermitian open systems},
  journal = {Nat. Rev. Phys.},
  volume  = {6},
  pages   = {11--27},
  year    = {2024},
  doi     = {10.1038/s42254-023-00659-z}
}

@article{Huang2025MergingBICs,
  author  = {Lujun Huang and Bin Jia and Artem S. Pilipchuk and Sibo Huang and Chen Shen and Almas F. Sadreev and Yong Li and Andrey E. Miroshnichenko},
  title   = {Merging bound states in the continuum in an open acoustic resonator},
  journal = {Sci. China Phys. Mech. Astron.},
  volume  = {68},
  pages   = {214311},
  year    = {2025},
  doi     = {10.1007/s11433-024-2496-9}
}

@article{Song2026TripleResonatorBIC,
  author  = {Chao Song and Quansen Wang and Han Jia and Zhiwei Guo and Lujun Huang and Yong Li},
  title   = {Anomalously robust Fabry-P{\'e}rot bound states in the continuum in a coupled triple-resonator system},
  journal = {Phys. Rev. B},
  volume  = {114},
  pages   = {L020201},
  year    = {2026},
  doi     = {10.1103/z7pd-hvsg}
}

@article{Wang2019AcousticEP,
  author  = {Xu Wang and Xinsheng Fang and Dongxing Mao and Yun Jing and Yong Li},
  title   = {Extremely Asymmetrical Acoustic Metasurface Mirror at the Exceptional Point},
  journal = {Phys. Rev. Lett.},
  volume  = {123},
  pages   = {214302},
  year    = {2019},
  doi     = {10.1103/PhysRevLett.123.214302}
}

@article{Zhu2018TopologicalEP,
  author  = {Weiwei Zhu and Xinsheng Fang and Dongting Li and Yong Sun and Yong Li and Yun Jing and Hong Chen},
  title   = {Simultaneous Observation of a Topological Edge State and Exceptional Point in an Open and Non-Hermitian Acoustic System},
  journal = {Phys. Rev. Lett.},
  volume  = {121},
  pages   = {124501},
  year    = {2018},
  doi     = {10.1103/PhysRevLett.121.124501}
}

@article{qj87-5xz9,
  title = {Robustness of bound states in the continuum in bilayer structures against symmetry breaking},
  author = {Semushev, Kliment V. and Zhao, Zilong and Proskurin, Alexey and Song, Mingzhao and Liu, Xinrui and Rybin, Mikhail V. and Maslova, Ekaterina E. and Bogdanov, Andrey A.},
  journal = {Phys. Rev. Appl.},
  volume = {25},
  issue = {1},
  pages = {014038},
  numpages = {13},
  year = {2026},
  month = {Jan},
  publisher = {American Physical Society},
  doi = {10.1103/qj87-5xz9},
  url = {https://link.aps.org/doi/10.1103/qj87-5xz9}
}
\end{document}